\documentclass{aa} 
\usepackage{graphicx}
\usepackage{auto-pst-pdf}
\newcommand{\Ni}{{$^{56}$Ni}}
\newcommand{\Nimass}{{M(^{56}\mathrm{Ni})}}
\usepackage{color}

\usepackage{footnote}
\makesavenoteenv{tabular}
\usepackage[toc,page]{appendix}
\begin{document} 
\title{A systematic study of magnetar-powered hydrogen-rich supernovae}
\author{Mariana Orellana\inst{1}\thanks{Member of the Carrera del Investigador Cient\'{\i}fico, CONICET, Argentina}, %
Melina C. Bersten\inst{2,3,4}, %
Takashi J. Moriya\inst{5}}

\institute{Sede Andina, Universidad Nacional de R\'{\i}o Negro, Mitre 630 (8400) Bariloche, Argentina
\and  Instituto de Astrof\'\i sica de La Plata (IALP), CCT-CONICET-UNLP. Paseo del Bosque S/N (B1900FWA), La Plata, Argentina
\and  Kavli Institute for the Physics and Mathematics of   the Universe (WPI), The University of  Tokyo, 5-1-5 Kashiwanoha, Kashiwa, Chiba 277-8583, Japan
\and  Facultad de Ciencias Astron\'omicas y Geof\'{\i}sicas, Universidad Nacional de La Plata, Paseo del Bosque S/N, B1900FWA La Plata, Argentina
\and  Division of Theoretical Astronomy, National Astronomical Observatory of Japan, National Institutes of Natural Sciences, 2-21-1 Osawa, Mitaka, Tokyo 181-8588, Japan}
\offprints{M. Orellana\\  \email{morellana@unrn.edu.ar}}

\titlerunning{H-rich magnetar SNe}

\authorrunning{Orellana et al.}
 \date{Accepted 31/08/2018}
\abstract {It has been suggested that some supernovae (SNe) may be powered by a magnetar formed at the moment of the explosion. While this scenario has mostly been applied to hydrogen-free events, it may be possible also for hydrogen-rich objects.}
{We explore the effect of including a magnetar on the light curves of supernovae with H-rich progenitors.}
{We have applied a version of our one-dimensional LTE radiation hydrodynamics code that takes into account the relativistic motion of the ejecta caused by the extra energy provided by the magnetar.
For a fixed red supergiant (RSG) progenitor, we have obtained a set of light curves that corresponds to different values of the magnetar initial rotation energy and the spin-down timescale. The model is applied to SN~2004em and OGLE-2014-SN-073, two peculiar Type II SNe with long-rising SN~1987A-like light curves, although with much larger luminosities.}
{The presence of a plateau phase in either normal or superluminous supernovae is one possible outcome, even if a magnetar is continuously injecting energy into the ejecta.
In other cases, the light curve shows a peak but not a plateau. Also, there are intermediate events with a first peak followed by a slow decline and a late break of the declining slope. Our models show that bright and long rising morphologies are possible even assuming RSG structures.}
{A large number of supernova discoveries per year reveal unexpected new types of explosions. According to our results, SLSNe II-P are to be expected, as well as a variety of light curve morphologies that can all be possible signs of a newly born magnetar.
}
\maketitle
\section{Introduction} 
\label{sec:intro}

Nowadays supernovae (SNe) are known to be varied phenomena. Their classification has evolved in parallel to the increasing amount of photometric and spectroscopic data and the long standing efforts to explain their nature. Superluminous SNe (SLSNe), discovered already more than a decade ago, reach 10 to 100 times larger luminosities than regular SNe \citep{Q11,G12}.
They present stimulating cases to explore the extents of the available theoretical models. One of the main proposals to provide the extra source powering the luminosity of SLSNe is the formation of a magnetar. The rotational energy of the hypothetical magnetar would be responsible for the extra energy needed to power the very bright light curve (LC). Although the magnetar model has been used previously in the literature (see e.g. \cite{2007Maeda} for the peculiar SN~2005bf), it became more popular after the works of \cite{W10} and \cite{KB10}. These authors showed that, if a rapidly rotating (millisecond period initially) neutron star with a large magnetic field ($B\sim 10^{14}$ G) is assumed to fully deposit its energy in the ejecta, the resulting SN may reach a peak luminosity in excess of $\simeq 10^{44}$ erg s$^{-1}$.
After this suggestion, the magnetar model was extensively used in the literature to explain several observed SNe.
In particular, the semi-analytic prescription by \cite{KB10} has become common thanks to its relative success to reproduce the morphology of the LCs (see e.g. \citealt{inserra2013}). Yet, such a treatment neglects both the formation and the expansion of the shock wave. We refer to \cite{2017Yu} or \citet{nicholl2017} for recent statistical studies applying this simple model. 

\cite{2016Suk} discussed the upper bounds to the energy that can be radiated by the different scenarios invoked in the literature. The most extreme cases can be explained by magnetars, though the details of the interaction between this energy reservoir and the rest of the stellar structure are not well established at the scale of the neutron star. Magnetar power has been mainly proposed as a possible central source for H-free SLSNe (or SLSNe~I), while interaction with the circumstellar medium (CSM) is the preferred model to explain hydrogen-rich SLSNe (or SLSNe~II) \citep{csm2011,moriya2013}. The reason for this is that most observed SLSNe~II are Type~IIn, i.e. objects that show narrow lines in their spectra, which is indicative of interaction between the SN ejecta and a dense CSM, \citep[e.g. as in SN~2006gy][]{2007Smith}. 
 However, there are a few cases lacking the narrow and intermediate-width line emission, such as SN~2008es, which was an H-rich non-Type IIn (\citealt{2009Miller,gezari2009}, see also \citealt{Inserra2016}). 
Some other SLSNe were initially H-poor but H$\alpha$ emission was later found \citep{Yan}. 
It could be possible that some of these objects were powered by a magnetar source. 
In addition to SLSNe~I, magnetar models have been used to explain other peculiar objects, such as the unusual SN~2005bf \citep{2006Folatelli,2007Maeda} and its recent analog SN PTF11mnb \citep{2017Taddia}.

Many efforts have been done to deal with magnetism in SN explosions \citep{2009Hu}. Current knowledge indicates that progenitors with fast-rotating iron cores likely develop magnetorotational instabilities \citep[e.g.][]{Akiyama,Heger} as part of the mechanism that increases the magnetic field strength \citep{Moesta15}. Simulations suggest that magneto-rotational explosions could be asymmetric. \cite{2007Burrows} have analyzed the dynamical effects of magnetic stresses on the SN, along with the possible jet formation that connects SNe with Gamma-Ray Bursts \cite[see also][]{Wheeler}. Recent works also discussed that eventual jets launched at the birth of the magnetar cannot be ignored during the explosion itself nor later when fall-back mass accretion might occur \citep{Soker17}. SN explosions might be asymmetric when influenced by a powerful magnetar. In that case the 1D approach is certainly unrealistic. \cite{2016Chen} studied the dynamical effect of the magnetar energy deposition based on 2D 
simulations. Although radiation transport is neglected, that work shows that fluid instabilities cause strong mixing and fracture shells of ejecta into filamentary structures which could affect photon emission. There are many issues related to the formation and deposition of the magnetar energy that remain unclear and that are beyond the scope of the present study.

In the context of H-rich progenitors, magnetar-powered LCs have not been deeply studied in the literature. \cite{2016BAAA} presented a tentative simulation for an RSG progenitor showing that, as expected for this type of progenitor, the plateau phase is still present in some cases when a magnetar source is taken into account. More recently, \cite{ST2017} and \cite{DA17} discussed similar scenarios, the former focused on on ordinary Type~II-P SNe and the latter on SLSNe. In this work we analyze whether both cases can be embraced by variations of the magnetar characteristics. 

The rest of this paper is organized as follows. Our calculations are performed with the code described in \citet{Bersten11}. In section~\ref{sec:model} we explain the modifications that we introduced in the code in order to treat this problem.
The effect of magnetar parameters on the LC shape is discussed in section~\ref{sec:parameters}, where we present our systematic analysis as a natural extension of our previous studies. In section~\ref{sec:comparison} we  apply this model to the peculiar SN 1987A-like bright SN OGLE--2014--SN--073 (hereafter OGLE14-073), recently published by \cite{Terreran}. The H-rich magnetar model can be applied to explain this interesting source that is one of the brightest SNe II ever discovered. We also devote a tentative parameter exploration applied to SN~2004em, another peculiar SN~1987A-like object. Discussion and conclusions including comparisons with previous works are presented in section~\ref{sec:conclusion}.
 
\section{Numerical model}
\label{sec:model}

The inclusion of a magnetar source in our one-dimensional hydrodynamical code was recently implemented in \cite{Bersten16}. The main difference in the current work is the progenitor structure used as initial condition of the calculations. We are now interested in analyzing the possible effect of a magnetar in H-rich objects. Therefore, we assume a red supergiant structure with a thick H-envelope, typical of Type~II-P SN progenitors. Our code self-consistently follows the whole evolution of the SN explosion starting from a given pre-SN structure in hydrostatic equilibrium, i.e., the shock wave propagation in the stellar interior, the shock breakout, and the subsequent expansion phases. 
The explosion is simulated by artificially injecting thermal energy near the center of the progenitor star, without specification of the involved mechanism. A few seconds later, after the neutron star (NS) is already formed, an extra source of energy due to the magnetar is incorporated. 
The code assumes flux-limited radiation diffusion for optical photons and a one-group approximation for the non-local deposition of gamma-rays produced by radioactive decay of \Ni\, \citep[for more details see][]{Bersten11}.

To parameterize the magnetar source we use a spin-down timescale ($t_p$) and an initial rotation energy ($E_{\rm rot}$) as the free parameters of the model. They enter into the basic expression for the energy supplied per unit time by the magnetar as

\begin{equation}
L(t)= \frac{E_{\rm rot}}{t_p} \bigg(1+\frac{t}{t_{p}}\bigg)^{-2}.
\label{magnetar}
\end{equation}

These alternative parameters are equivalent to the usual $B$ (magnetic field) and $P$ (initial rotation period), but in this way we avoid to include explicit properties of the NS, such as the radius or the moment of inertia, which might be afterwards explored by assuming a specific equation of state \citep[see][for more details]{Bersten16}. Although the presence of a strong magnetic field in the NS interior and its coupling with matter is not fully understood, studies of the cooling of magnetized NSs \citep[e.g.][]{Turolla15} have shown that the initial $B$ value is preserved for at least a few thousand years. Thus, the magnetars known today were born spinning very fast but with similar magnetic field to their current extreme value $B \ge 10^{13}$~Gauss.

Our strong assumption is that $L(t)$ is fully deposited and thermalized in the inner layers of the exploding star as a persistent energy injection. Specifically, we deposit the magnetar energy in the inner 15 zones of the progenitor model assuming a box function in mass coordinate.
Full deposition is usually assumed in the literature, although the option of inefficient heating by the nascent magnetar was explored by \cite{KMB2016} in order to obtain a double peaked LC. Also, the leakage of hard emission was discussed by \cite{2016W} as an interesting alternative to full energy trapping.

In our treatment, if the photosphere recedes deep enough so that magnetar energy is deposited at optically thin layers, we add the magnetar contribution to the bolometric luminosity. Although the power engine is located deep into the ejecta, its influence propagates outwards pushing the lightweight outer shells up to enormous velocities. In some cases, this can lead to relativistic movements, specially in extreme cases where the energy injected by the magnetar is several orders of magnitude larger than the explosion energy, as we showed in \cite{Bersten16}. Therefore, we have modified our code to take this effect into account. In the Appendix~A we present the formulation of the relativistic hydrodynamics included in our 1D code.

The pre-SN models adopted throughout this work were calculated by \cite{NH88} following the stellar evolution until core collapse. Specifically, RSG progenitors with masses of 15 and 25 $M_\odot$ are used in this study. These stellar models assume solar metallicity and no rotation. However, low-metallicity and rotating stars are probably more realistic progenitors of rapidly rotating and strongly magnetized NSs than our pre-SN models. Although this is a caveat in our analysis, we note that magnetism and rotation in massive stars are complex problems for which there is still no definitive solution \citep{Heger2003,Heger}.

\section{Exploration of the parameter space} 
\label{sec:parameters}

Preliminary results of the magnetar effects in H-rich progenitors were presented in \cite{2016BAAA}. That study clearly shows that the plateau morphology of the LCs can be preserved in some cases. 

In this section we consider a fixed progenitor star with main-sequence mass of 15~$M_{\odot}$, pre-explosion radius of 500~$R_\odot$, and surface metallicity of $Z\sim 0.02$. This pre-SN model shows a transition between H-rich to He-rich layers at $\approx 3.2 M_\odot$. More details on chemical abundances can be seen in Appendix~\ref{appendix:progenitor}, and a summary table with futher progenitor properties is later presented in \S~\ref{sec:conclusion}. First, we discuss this reference model (\S~\ref{sec:onemodel}) and then we focus on a grid of models (\S~\ref{sec:grid}).

\subsection{Comparison: model with and without magnetar}
\label{sec:onemodel}
\begin{figure}
\resizebox{\hsize}{!}{\includegraphics[angle=0]{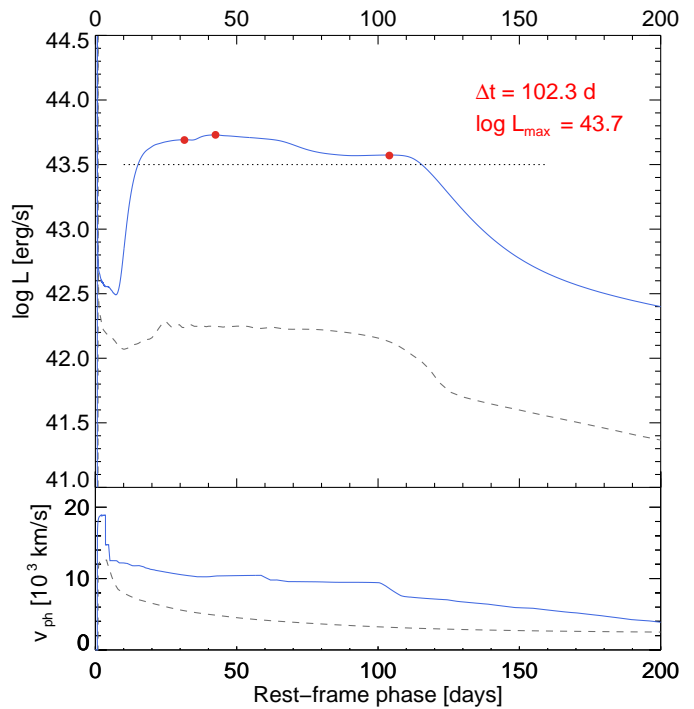}}
\caption{Light curve (top panel) and photospheric velocities (bottom panel) for our reference model (see text) shown in blue solid line, corresponding to a magnetar with $E_{\rm rot}=10$~foe and $t_p=1$~d. A characteristic maximum luminosity $L_{\rm max}$ is derived as the mean of the three local maxima found in the LC (red dots). The intersection of the horizontal line defined as $\log L_{\rm max} - 0.2$~dex, with the LC provides the estimated temporal extent of that maximum, $\Delta t$. For comparison, we show in dashed gray lines the same SN model without a magnetar.} 
\label{fig:definition}
\end{figure}

Figure~\ref{fig:definition} shows a comparison between models with and without a magnetar source for the progenitor star described above. We further adopted an explosion energy of $1.5$ foe (1 foe $= 1 \times 10^{51}$ erg) and a \Ni\, mass of $0.1 M_\odot$. For the magnetar source, values of $E_{\rm rot}=10$~foe and $t_p= 1$~d were used. It is clear that in the presence of a magnetar, the plateau luminosity and duration can change substantially. Also, differences in the phospheric velocity evolution are notable. Models with magnetars produce higher velocities. An interesting feature of magnetar models is the existence of a short phase of increasing luminosity preceding the plateau phase. This rise can be as large as one order of magnitude, which is much greater and steeper than in the case without a magnetar. This feature of the magnetar models can help to distinguish the power source of the SN if it is discovered early enough.

\begin{figure*}
\includegraphics[width=0.49\hsize,angle=0]{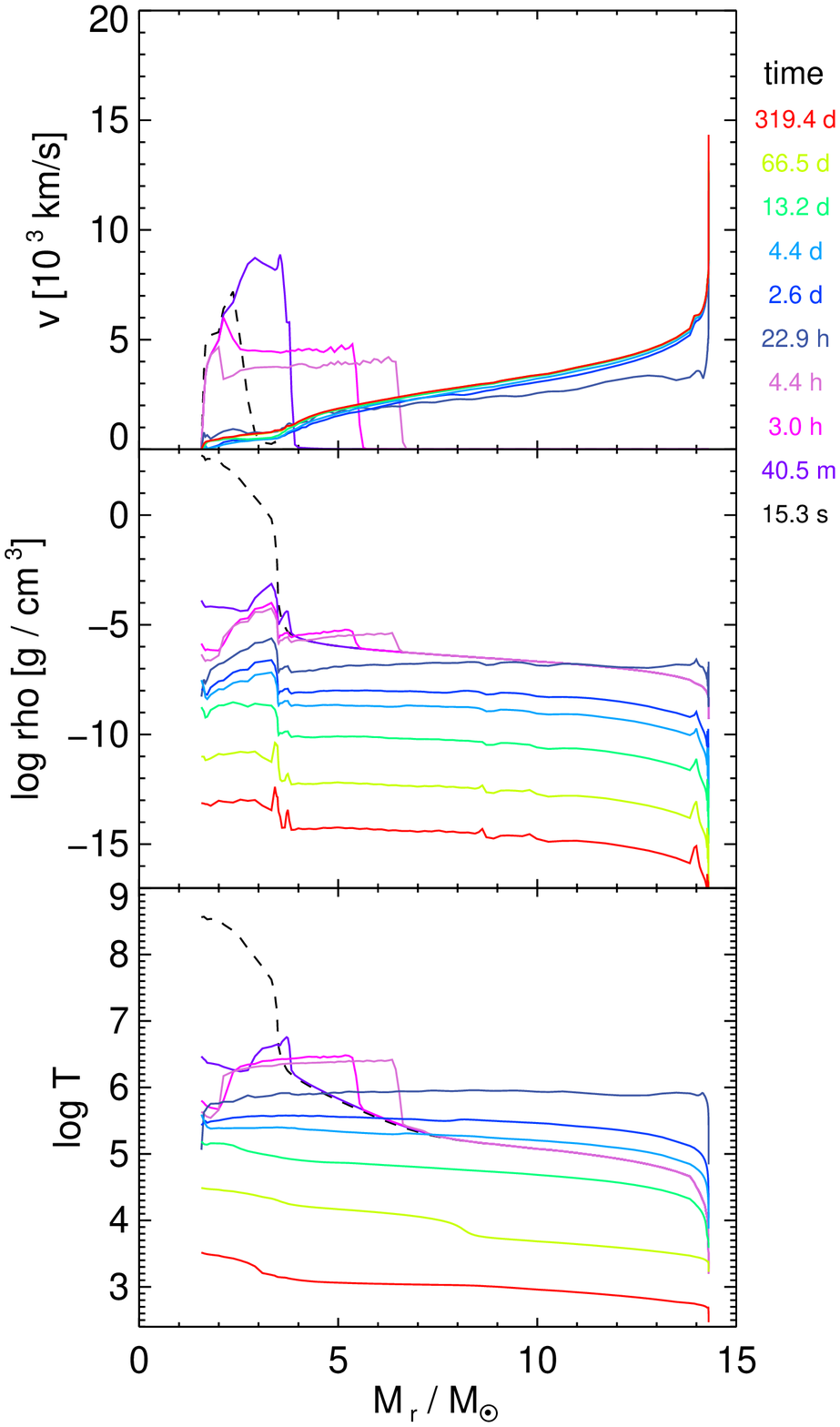}\hfill%
\includegraphics[width=0.49\hsize,angle=0]{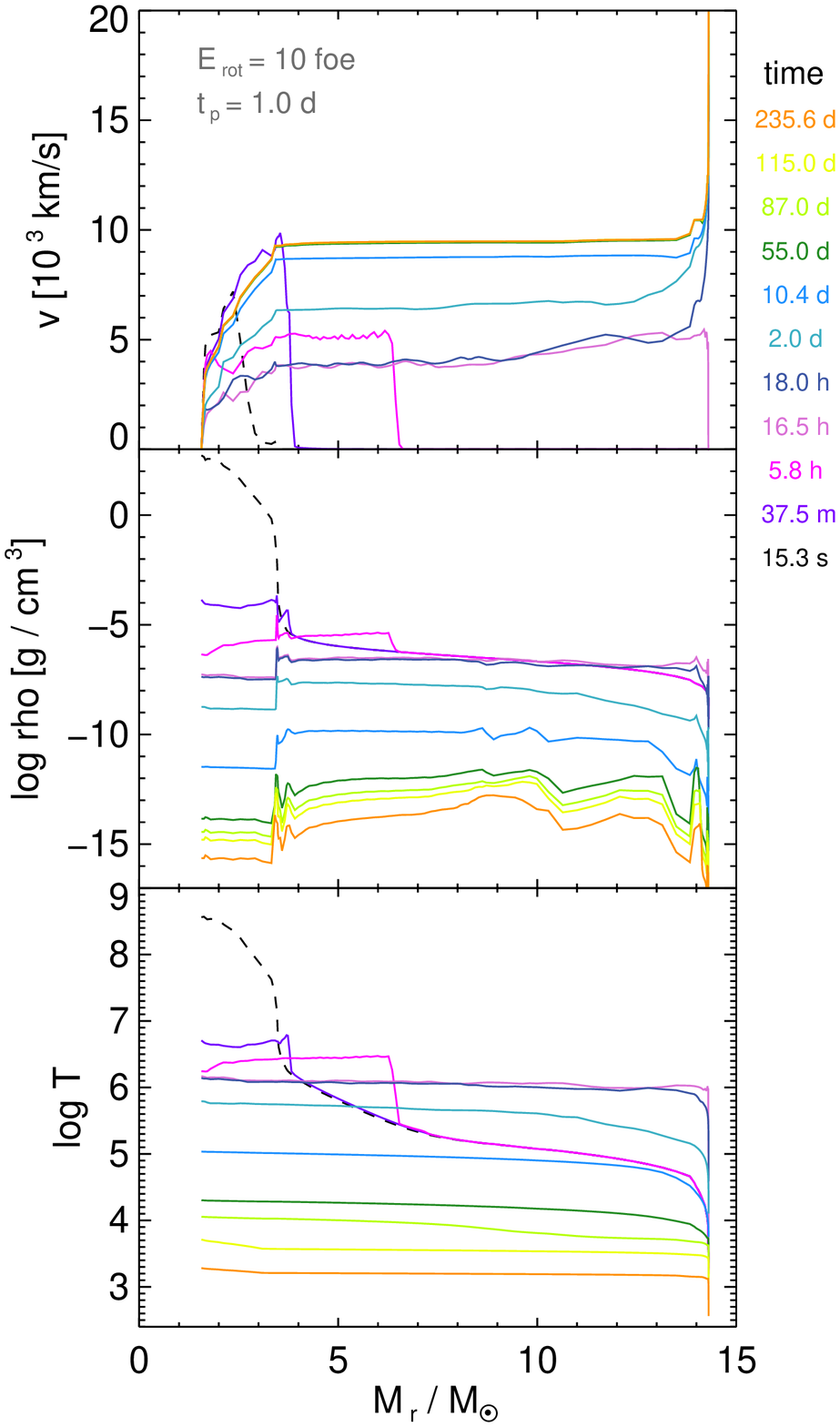}\\
\caption{Effect of the magnetar on the velocity, density and temperature profiles, as a function of the Lagrangian mass coordinate. The panels on the right show the same pre-SN model as on the left but including a magnetar with $E_{\rm rot}=10$~foe and $t_p=1$~d, and all other parameters fixed. 
The color code of the epochs is preserved in each column, and is measured starting at the moment of $E_k$ injection.
With this powerful magnetar (right panels) the maximum velocity reaches one third of the speed of light at the edge of the ejecta. We cut the velocity axis in order to show the most relevant interval.}
\label{fig:profiles} 
\end{figure*} 

\begin{figure}
\resizebox{\hsize}{!}{\includegraphics[angle=-90]{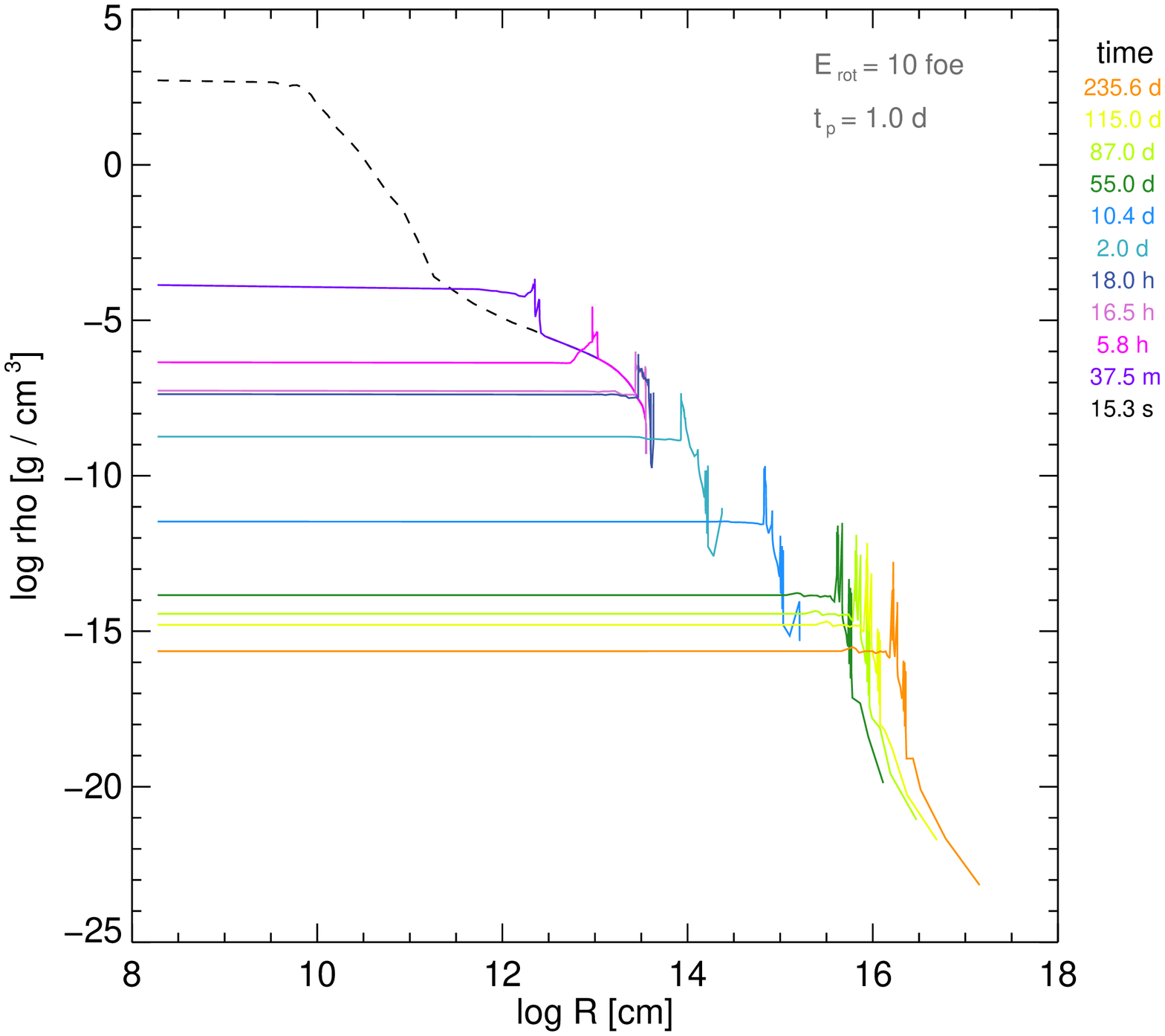}}
\caption{Density profiles as function of the radius for the same magnetar of Figure~\ref{fig:definition} and epochs as in Figure~\ref{fig:profiles}.} 
\label{fig:rho_R}
\end{figure} 

For a deeper comparison between the models presented in Figure~\ref{fig:definition}, the profiles of different physical quantities (velocity, density and temperature) at some specific epochs after the explosion are shown in Figure~\ref{fig:profiles}. The most notable differences are in the velocity profiles. Almost the entire ejecta reach very fast velocities in the presence of a magnetar. This explains the differences in the photospheric velocities seen in lower panel of Figure~\ref{fig:definition}. It is interesting to note that homologous expansion is reached around 4 days after the explosion for the model without a magnetar whereas it is delayed until around 50 days for the magnetar model. This means that the ejecta dynamics is modified after the shock break-out by the the extra magnetar-powered force. As a result, the inner density of the ejecta becomes extremely low at the final phases of the simulation, while most of the ejected matter $(\sim 10\; M_\sun)$ moves with speed $\sim 10^4$~km/s. 
Figure~\ref{fig:rho_R} shows the radial distribution of the mass density into the ejecta. 
Note that a thin denser shell is formed as the supernova expands without any opposing
pressure outside, therefore starting to create a large bubble.
The overall behavior found is consistent with the 2D-simulations by \cite{2016Chen} who pointed out that instabilities arise from the piling up of radiatively accelerated matter. This indicate that the full trapping of the magnetar power and the 1D approach may be questionable due to the unstable configurations produced. 

We define some quantitative parameters which can help to characterize and compare LC morphologies. We call $L_{\rm max}$ the mean value of the local maxima produced after the shock peak, as illustrated in Figure~\ref{fig:definition}. In some cases, only one clear maximum is obtained. To characterize the temporal extent of the LC we measure the interval $\Delta t$ over which $\log L > (\log L_{\rm max} - 0.2$ dex). The value of $0.2$~dex in our definition is motivated by \cite{tesis_Meli} and references therein.
These parameters are similar to the plateau luminosity and duration in the cases resembling Type II-P SNe.
In this regard, we note that there is a variety of similar quantities defined elsewhere in the literature.
A recent discussion on the duration--luminosity phase space of optical transients by \cite{Villar} applies a somewhat similar definition to ours, whereas the observational treatment proposed by \cite{Olivares2010} is only applicable if a plateau phase and a clear transition from the plateau to the decline tail\footnote{A transition is present in cases that we call intermediate but probably the nebular phase deserves a more careful treatment than our simplified analysis.} can be traced. In the investigation of magnetar-powered ordinary Type II-P SNe \cite{ST2017}, the plateau duration is measured in a very different way as the time from the explosion until the moment when the photospheric radius falls below $10^{14}$~cm. This definition, although useful from the theoretical point of view, is not directly measurable in observations. In the next section we measure the parameters defined here ($L_{\rm max}$, $\Delta t$) for a set of magnetar parameters.

\subsection{Grid of models}
\label{sec:grid}
We have calculated a set of SN LC models for different values of $t_p$ and $E_{\rm rot}$. The considered values are intended to cover as much of the parameters range as possible for the progenitor previously described. We have assumed a standard explosion energy ($E_k =1.5 \times$ 10$^{51}$ erg) and \Ni\ production ($0.1$~$M_{\odot}$). In Table~\ref{models} we provide information of the magnetar parameters for the grid of simulations performed here, together with the LC parameters as defined in the previous seccion ($L_{\rm max}$ and $\Delta t$). \cite{M15} demonstrate that the maximum available rotational energy (without accounting for gravitational waves) of a NS is in the range of 90 -- 165 foe. Here we restrict to $E_{\rm rot}= 1 - 100$~foe, and $t_p = 0.03 - 10$ days.
The two most extreme conditions, $t_p =10$ days and $E_{\rm rot} = 30$ and 100 foe, respectively, were not capable to run with the same configuration as the others, so we choose not to include them here.

\begin{center}
\begin{table}
\caption[]{Model parameters and main characteristics of the magnetar (period, magnetic field strength) for the 15 $M_{\odot}$ RSG stellar progenitor. Each model has fixed $t_p$ and $E_{\rm rot}$. The observables from the light curve ($L_{\rm max}$ and $\Delta t$) result from our numerical simulations.}
\begin{tabular}{ccccccc}
\hline
\hline\noalign{\smallskip}
mod.& $t_p$ & $E_{\rm rot}$ & P $^\dag$& B $^\dag$&log $L_{\rm max}$&$\Delta t$\\[0.001cm]
              & [d] &[foe] & [ms] & [$10^{14}$ G] & [erg s$^{-1}$] & [d]\\
\noalign{\smallskip}\hline\noalign{\smallskip}
       0 & -- & -- & -- & -- & 42.25 & 100.3 \\
       1 & 0.03 & 1.0 & 5.07 & 72.60 & 42.47 & 87.8 \\
       2 & 0.03 & 3.0 & 2.92 & 41.91 & 42.73 & 73.8 \\
       3 & 0.03 & 10.0 & 1.60 & 22.96 & 43.15 & 57.5 \\
       4 & 0.03 & 30.0 & 0.92 & 13.25 & 43.57 & 52.1 \\
       5 & 0.03 & 100.0 & 0.51 & 7.26 & 43.99 & 38.5 \\
       6 & 0.1 & 1.0 & 5.07 & 39.76 & 42.51 & 83.8 \\
       7 & 0.1 & 3.0 & 2.92 & 22.96 & 42.88 & 67.8 \\
       8 & 0.1 & 10.0 & 1.60 & 12.57 & 43.32 & 59.4 \\
       9 & 0.1 & 30.0 & 0.92 & 7.26 & 43.78 & 61.6 \\
      10 & 0.1 & 100.0 & 0.51 & 3.98 & 44.30 & 28.5 \\
      11 & 0.1 & 300.0 & 0.29 & 2.30 & 44.72 & 12.1 \\
      12 & 0.3 & 1.0 & 5.07 & 22.96 & 42.60 & 85.6 \\
      13 & 0.3 & 3.0 & 2.92 & 13.25 & 43.04 & 70.5 \\
      14 & 0.3 & 10.0 & 1.60 & 7.26 & 43.51 & 74.8 \\
      15 & 0.3 & 30.0 & 0.92 & 4.19 & 44.00 & 48.0 \\
      16 & 0.3 & 100.0 & 0.51 & 2.30 & 44.55 & 20.5 \\
      17 & 1.0 & 1.0 & 5.07 & 12.57 & 42.86 & 78.8 \\
      18 & 1.0 & 3.0 & 2.92 & 7.26 & 43.25 & 85.8 \\
      19 & 1.0 & 10.0 & 1.60 & 3.98 & 43.67 & 102.3 \\
      20 & 1.0 & 30.0 & 0.92 & 2.30 & 44.21 & 42.2 \\
      21 & 1.0 & 100.0 & 0.51 & 1.26 & 44.68 & 16.5 \\
      22 & 3.0 & 1.0 & 5.07 & 7.26 & 43.06 & 88.6 \\
      23 & 3.0 & 3.0 & 2.92 & 4.19 & 43.44 & 108.9 \\
      24 & 3.0 & 10.0 & 1.60 & 2.30 & 43.88 & 77.7 \\
      25 & 3.0 & 30.0 & 0.92 & 1.33 & 44.29 & 43.8 \\
      26 & 3.0 & 100.0 & 0.51 & 0.73 & 44.69 & 18.1 \\
      27 & 10.0 & 1.0 & 5.07 & 3.98 & 43.23 & 104.4 \\
      28 & 10.0 & 3.0 & 2.92 & 2.30 & 43.55 & 138.6 \\
      29 & 10.0 & 10.0 & 1.60 & 1.26 & 44.00 & 74.4 \\
      30 & 30.0 & 1.0 & 5.07 & 2.30 & 43.33 & 123.4 \\
      31 & 30.0 & 3.0 & 2.92 & 1.33 & 43.68 & 102.0 \\
\noalign{\smallskip}
\hline\\[0.01cm]
\multicolumn{7}{l} {\small $^\dag$ Assuming $I=1.3\times 10^{45}$~g~cm$^2$, and $R=10$~km for the NS.}\cr
\end{tabular}
  \label{models}
\end{table}
\end{center}

\begin{figure*}
\includegraphics[width=0.33\hsize,angle=0]{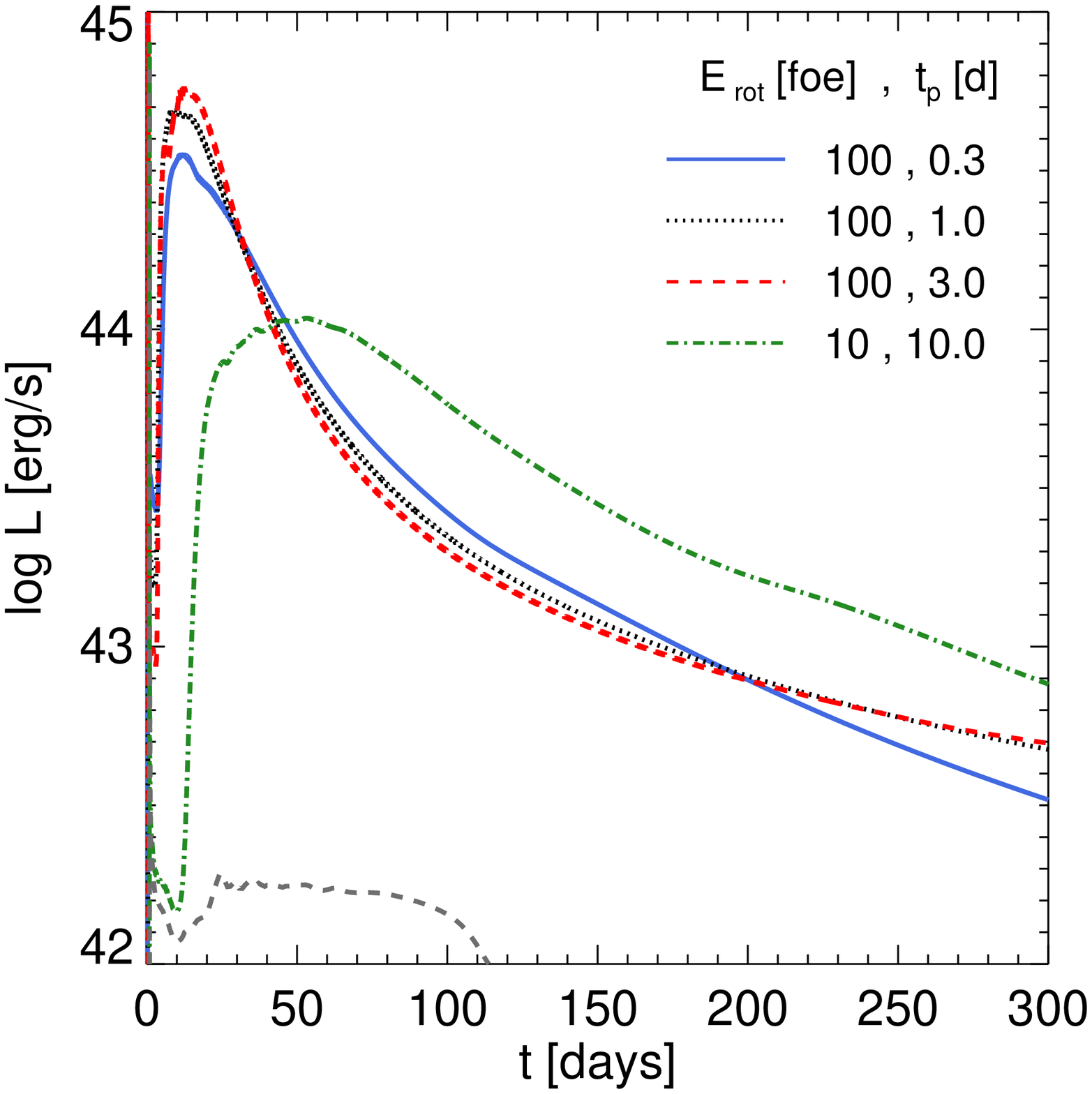}\hfill%
\includegraphics[width=0.33\hsize,angle=0]{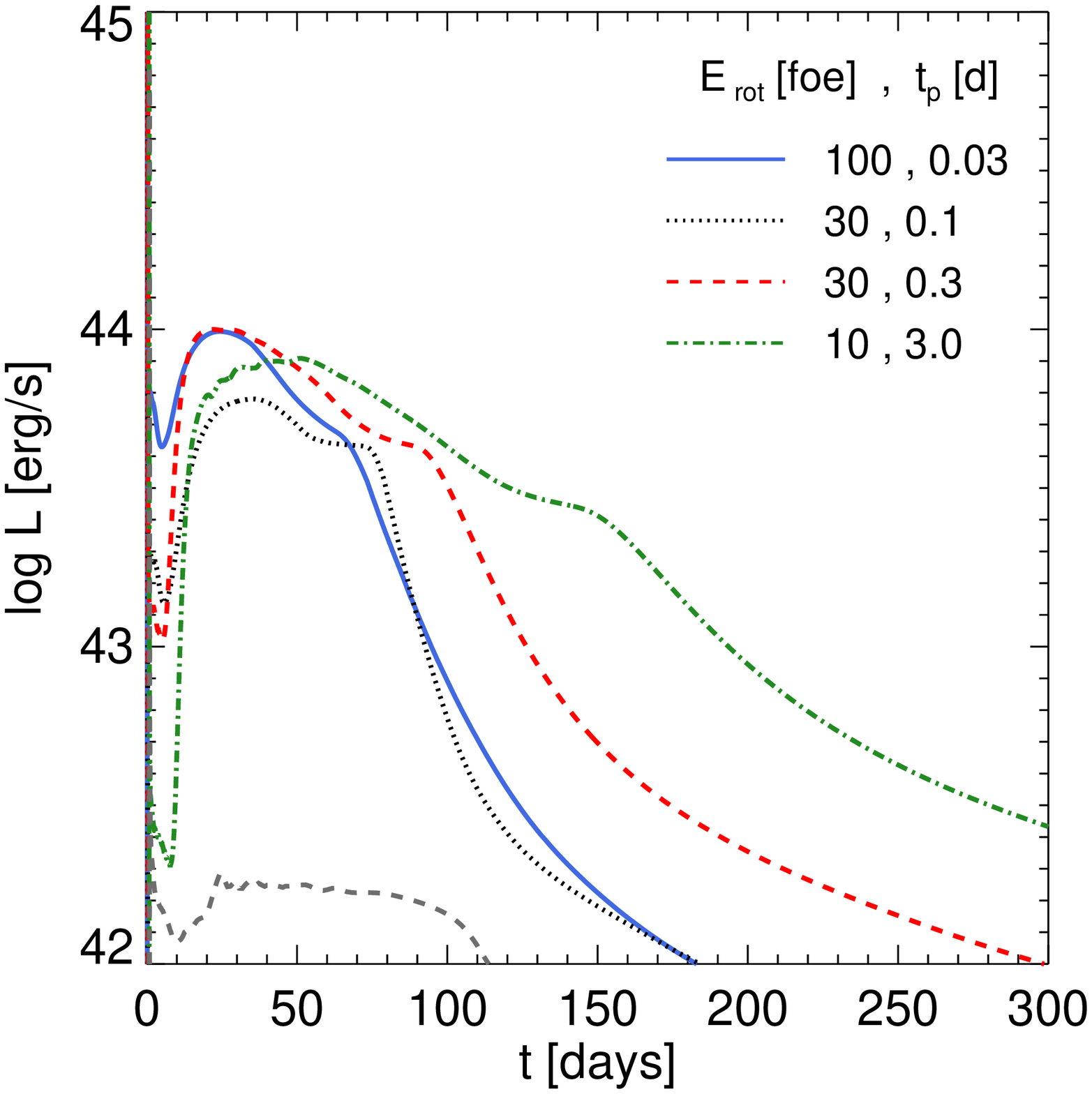}\hfill%
\includegraphics[width=0.33\hsize,angle=0]{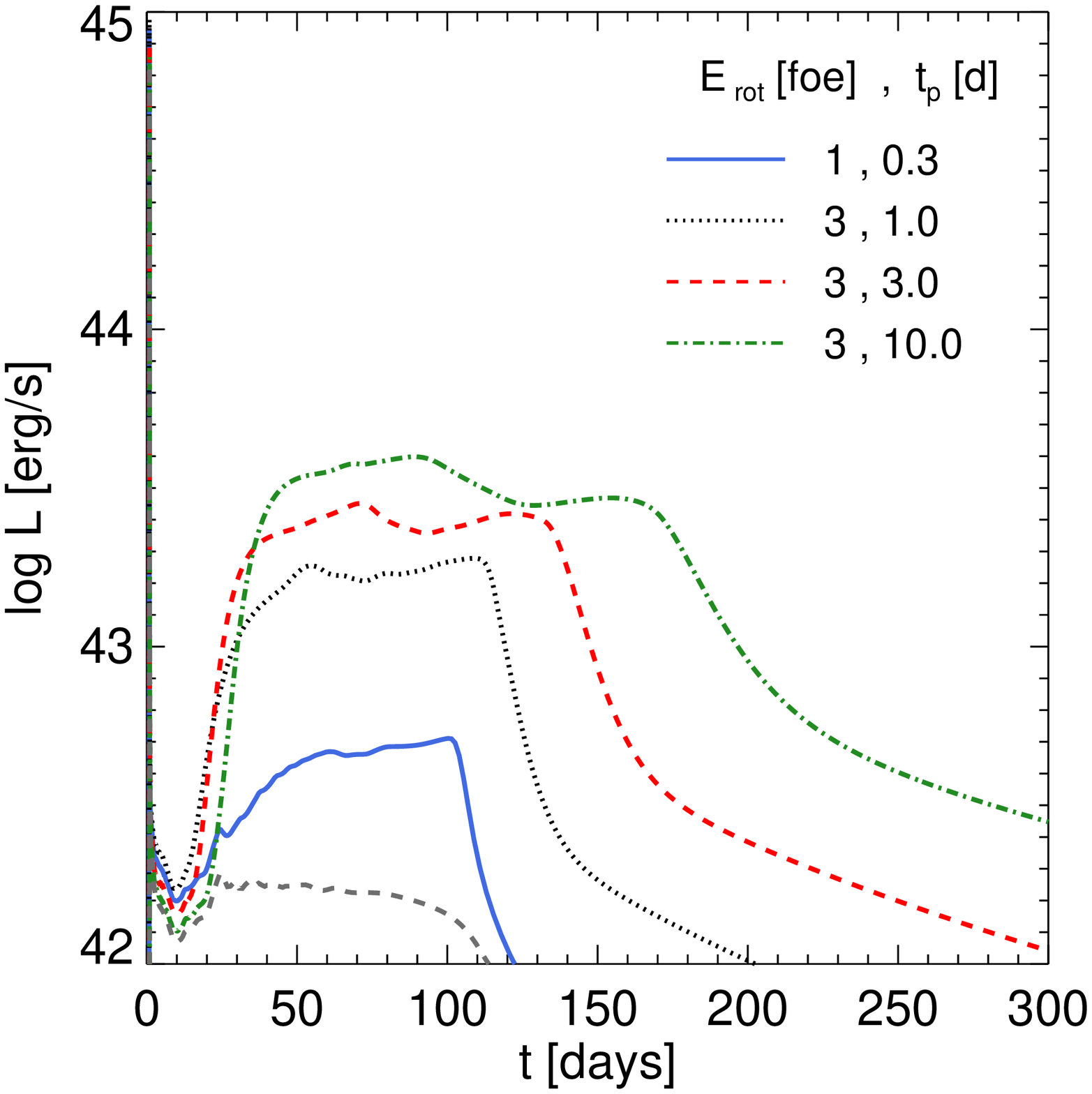}
\caption{Representative examples of the computed LC set. The left panel shows the LCs with one clear peak, the central panel shows slower declining LCs (or intermediate cases) usually presenting a broken evolution in the slope, and the right panel presents cases with a plateau, i.e. bright Type~II-P.
Legends indicate the parameters $E_{\rm rot}$ in units of $10^{51}$~erg, and $t_p$ in days. For comparison we include in dashed gray line the LC of the same SN without a magnetar.}
\label{fig:cases} 
\end{figure*}

Figure~\ref{fig:cases} illustrates the distinct kinds of LC morphologies that we obtained. Some LCs present a well-defined peak, while others show a plateau phase. There are intermediate cases that show a slow decrease after the peak and a later break in the decline slope at the transition to the tail. The late-time slope at $t > 200$ days is determined by the competing magnetar energy supply (Eq.~\ref{magnetar}) and the Ni--Co--Fe radioactive deposition power.

\begin{figure*}
\includegraphics[width=0.33\hsize,angle=0]{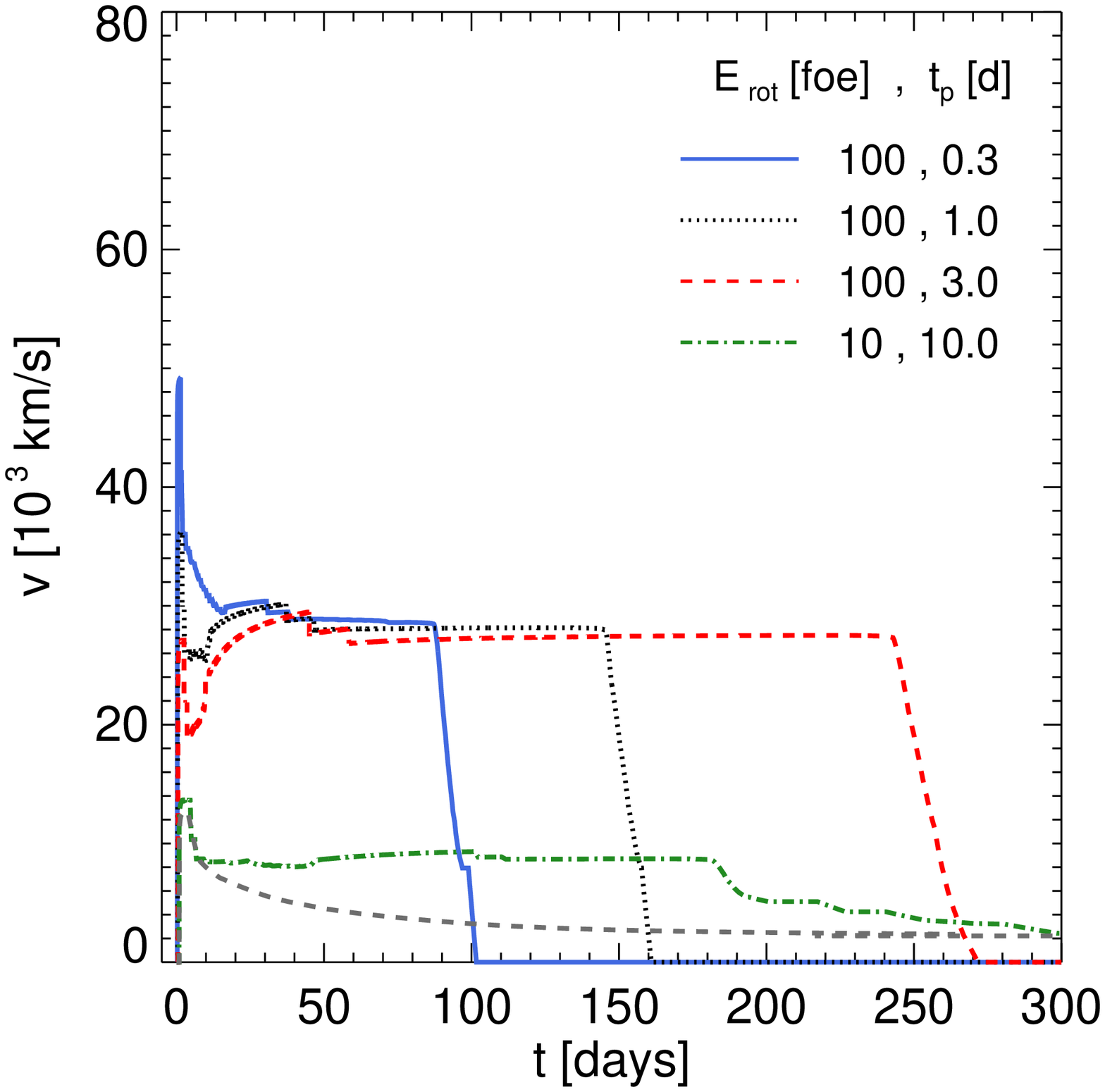}\hfill%
\includegraphics[width=0.33\hsize,angle=0]{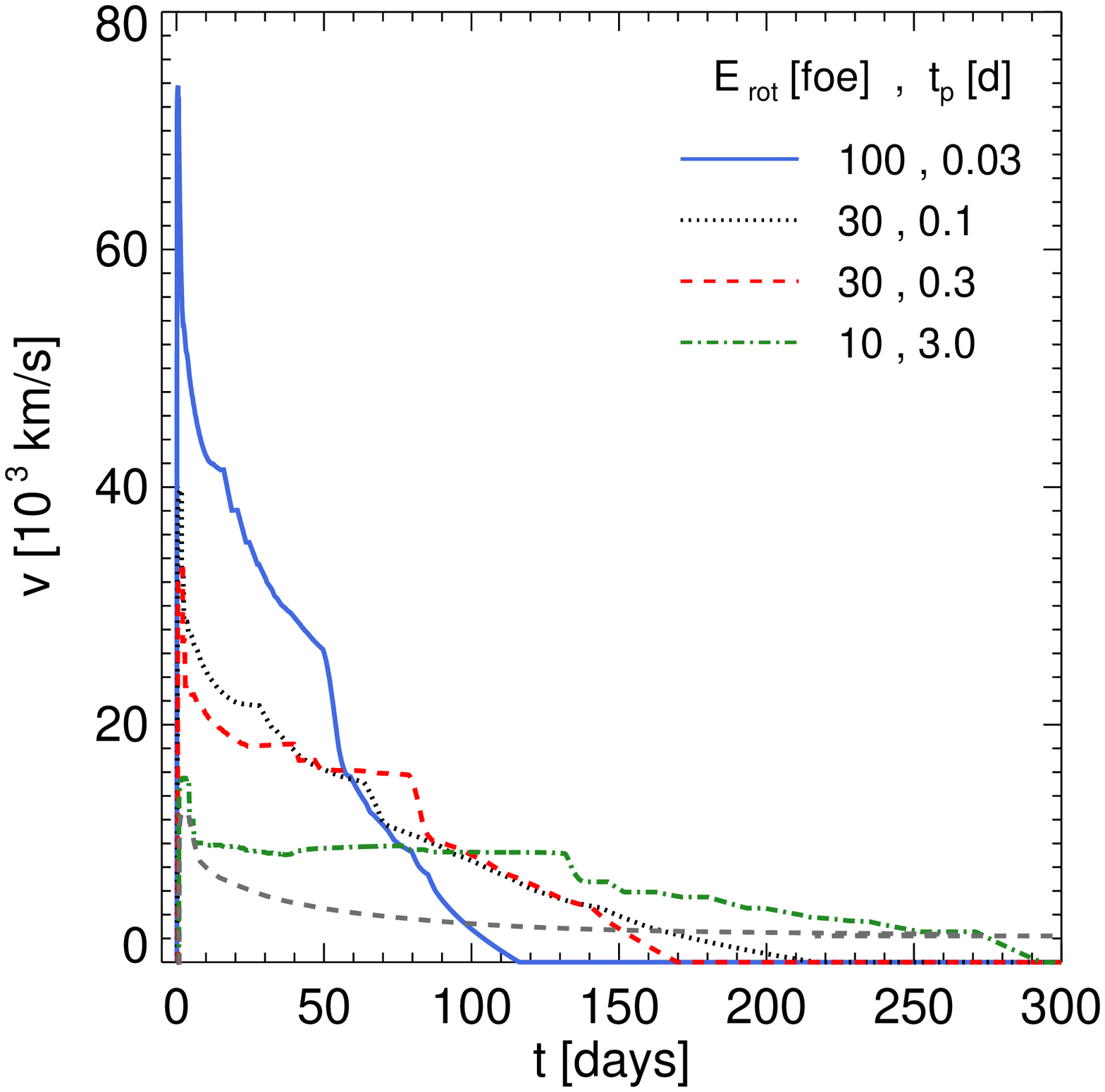}\hfill%
\includegraphics[width=0.33\hsize,angle=0]{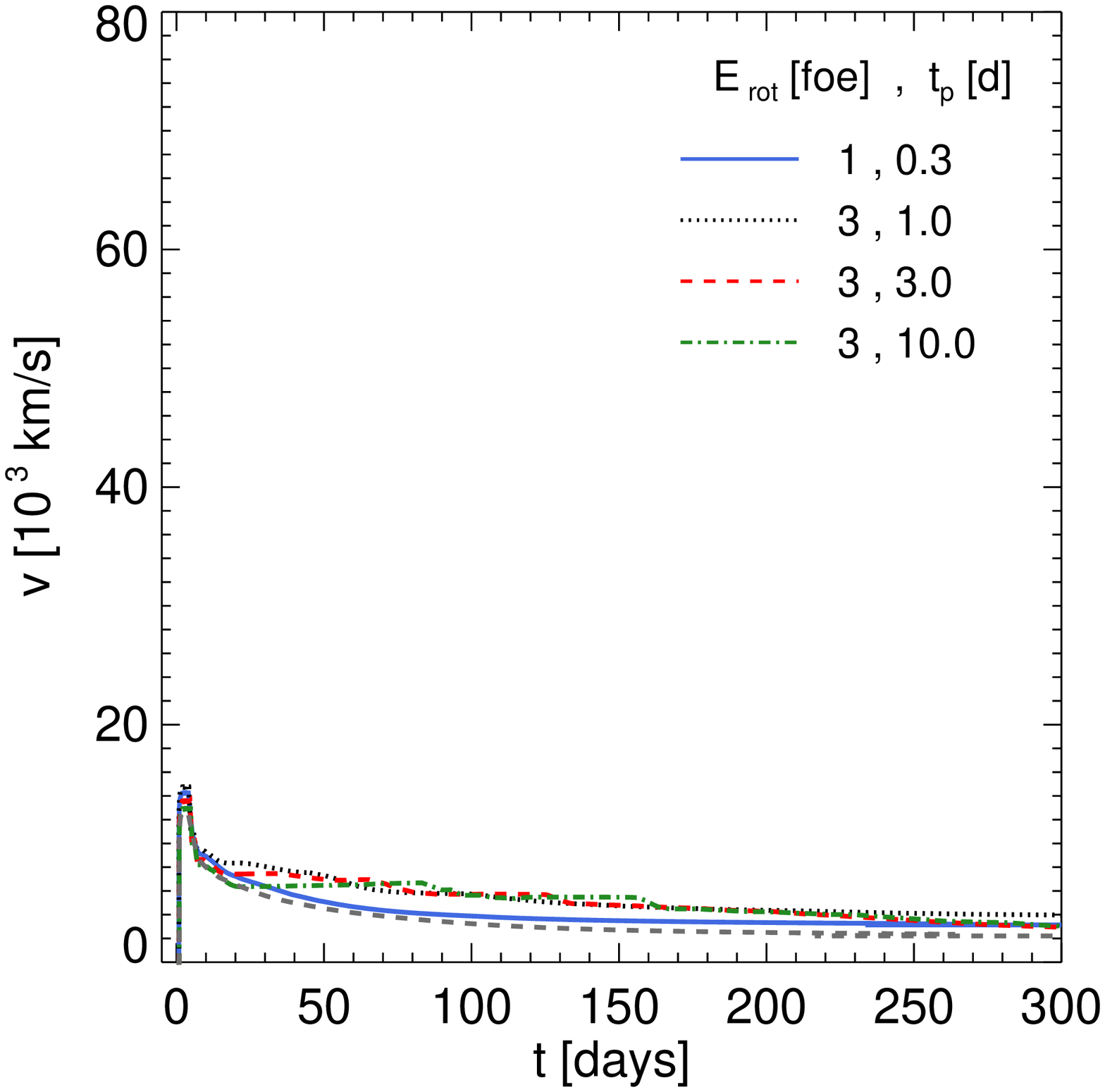}
\caption{Photospheric expansion velocity for the examples of Figure~\ref{fig:cases}. This velocity is null when the photosphere reaches the NS surface, i.e. when the entire ejecta becomes transparent. Legends are the same as in Figure~\ref{fig:cases}}
\label{fig:velo} 
\end{figure*} 

Regarding the expansion of the ejecta, in Figure~\ref{fig:velo} we present the model photospheric velocities. We note that the photospheric velocities seem to be more dependent on $E_{\rm rot}$ than on $t_{\rm p}$. However, this is not easy to connect with the kinetic energy of the ejecta due to the important effect of recombination on the photospheric velocities; although see \cite{2016W} for an alternative analytical treatment of the energetics. Our results show that larger values of $E_{\rm rot}$ produce larger photospheric velocities, i.e. a more important dynamical effect. In some cases, the expansion leads to an increase in the photospheric velocity during some time, as has been observed, for instance, in the peculiar SN~2005bf \citep{2006Folatelli}.

When considering the integrated luminosities during the whole SN evolution, models with low $E_{\rm rot}$ are more efficient in converting the magnetar energy into radiation. Models with $E_{\rm rot}=1$~foe can radiate up to a third of the energy injected by the magnetar, while for $E_{\rm rot}=100$~foe, this efficiency is $\lesssim 2\%$.

For low $E_{\rm rot}$, though the dynamics of the ejecta seems less affected (see Figure~\ref{fig:velo}) the velocities during the plateau phase are systematically larger than in the case without a magnetar.
With increasing $E_{\rm rot}$ more is energy available, thus the photosphere gets larger velocities at earlier times. Hence the ejecta dilute before, so the nebular phase might be reached earlier (left and medium panels of Figure~\ref{fig:velo}). For a fixed $E_{\rm rot}$, the photospheric velocities evolve faster for decreasing $t_p$.

\begin{figure}
\resizebox{\hsize}{!}{\includegraphics[scale=.50]{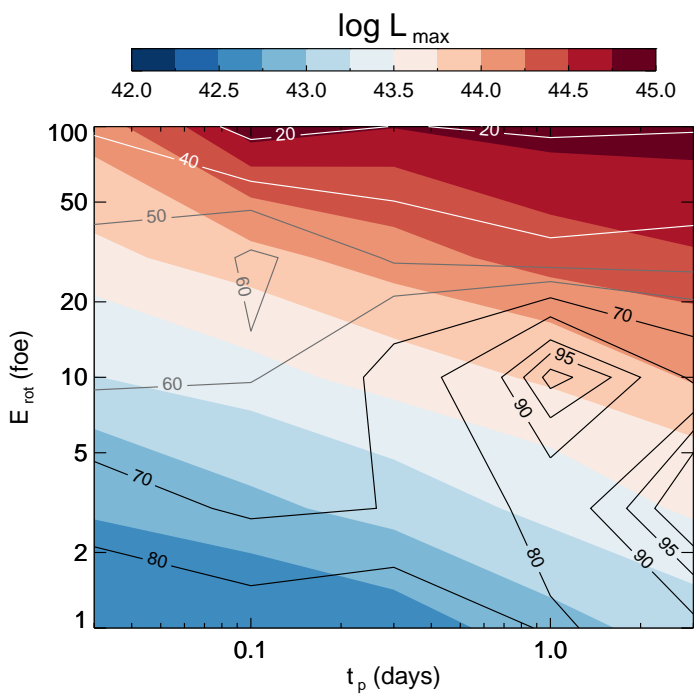}}
\caption{Main characteristics of the resulting LCs for a 15 $M_{\odot}$ RSG progenitor. The color scale indicates the maximum LC luminosity after the shock breakout. Contours show the estimated temporal extent of the maximum (lines of constant $\Delta t$, see Fig.~\ref{fig:definition}). The smallest, unlabeled contour corresponds to 100~days.} 
\label{fig:gridLplat}
\end{figure}
\begin{figure}
\resizebox{\hsize}{!}{\includegraphics[angle=-90]{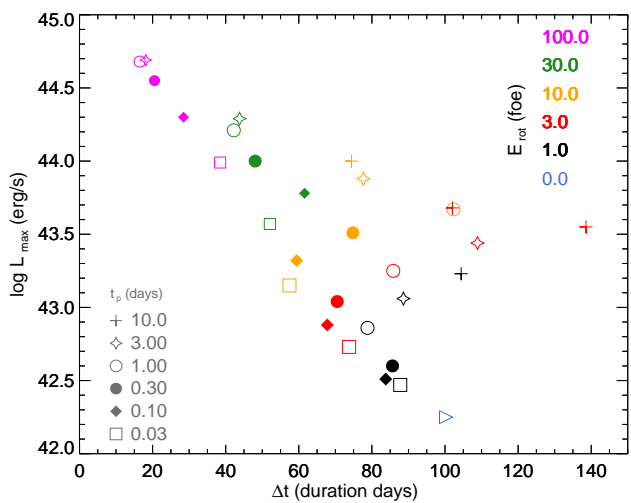}}
\caption{Luminosity versus duration of the LCs. Each modeled LC is characterized by the rotational energy (color code) and spin-down time scale (symbols) of the magnetar, whereas all the other characteristics are fixed (see text). The triangle presents the model without a magnetar.}
\label{fig:scatter} 
\end{figure}

LC observables such as $L_{\rm max}$ and $\Delta t$, and their relation with the magnetar parameters are shown in Figure~\ref{fig:gridLplat}. This figure can be used to obtain a rapid first guess of the magnetar parameters that may reproduce an observed SN, as well as to understand the dependence of some observables with the magnetar parameters. Note that a similar analysis was done by \cite{KB10} but for H-free progenitors and using a different parameterization of the magnetar properties. Interestingly, for the cases having a plateau-like LC without a single peak, the duration $\Delta t$ resembles the plateau duration in ordinary SNe II-P, with a mean value $\Delta t \sim 80$ days, and extending up to $\sim 140$ days. 
On the other extreme, a few of our single-peak LCs would be considered bright and rapidly evolving transients. 
These are usually the cases with very large $E_{\rm rot}$.
The cases with low values of $E_{\rm rot}$ have peak luminosities below $\simeq 10^{43}$ erg~s$^{-1}$ and would not be called SLSNe. This is more evident in Figure~\ref{fig:scatter}, which suggests that, for the parameter space sampled here, there is a correlation between $L_{\rm max}$ and $E_{\rm rot}$ with a scatter inversely proportional to $E_{\rm rot}$.
Figure~\ref{fig:scatter} is also useful to visualize the individual values of the parameters used in our exploration.

\section{Application to observed SNe}
\label{sec:comparison}
In order to test if magnetar-powered H-rich SNe are a viable explanation to some observed events, we have modeled the evolution of two peculiar H-rich SNe (OGLE14-073 and SN~2004em). Table~\ref{table:summary} shows the model parameters used to model the objets discussed in this section. 

\subsection{OGLE14-073}
The recently reported OGLE14-073 \citep{Terreran} at $z=0.1225$ presented a bright and very broad LC. Its spectra show prominent P-Cygni features of hydrogen but no sign of interaction with a CSM. The slow spectrophotometric evolution for OGLE14-073 is consistent with a classification as a peculiar Type~II event, similar to SN~1987A but much brighter. The explosion date of OGLE14-073 is not well constrained. Large values for the explosion energy ($\sim 12$ foe) and ejecta mass ($\sim 60$ $M_\odot$) as well as a rather large \Ni\ mass ($> 0.47 M_\odot$) need to be invoked in order to match the maximum luminosity and the late decline. The extreme values required to explain the properties of this object indicate that possibly another source is responsible for its brightness. \cite{Terreran} presented a magnetar as a viable explanation and discussed alternative scenarios for this event. Recently, \cite{DA17} performed a magnetar-powered modeling of this object. A discussion comparing this work and ours is 
presented in \S\ref{sec:comparison}. In a different proposal, \cite{2017Moriya} studied OGLE14-073 as a possible fallback accretion-powered SN following a failed explosion of a massive star.

\begin{figure}
\resizebox{\hsize}{!}{\includegraphics{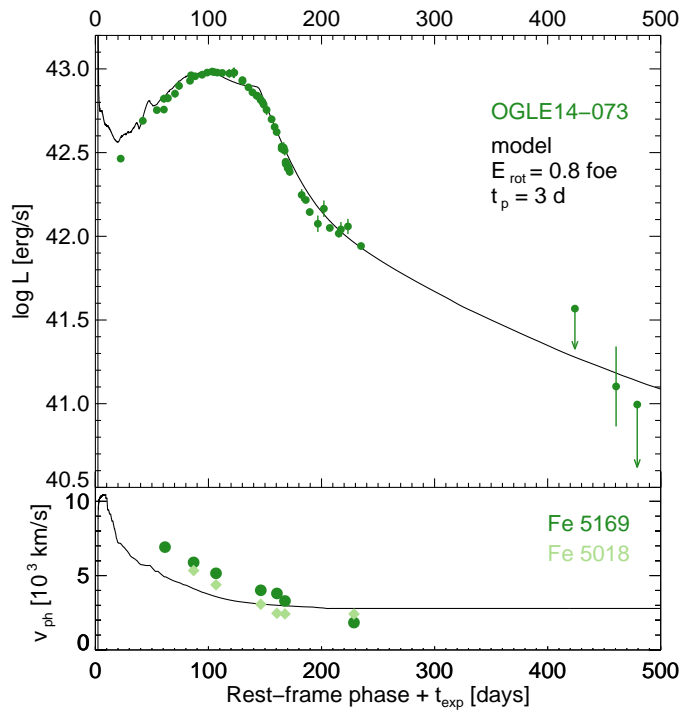}}
\caption{Magnetar-powered SN model for OGLE14-073. The SN is assumed to have exploded $t_{\rm exp}=20$~days before discovery. The points show the bolometric luminosities (upper panel) and Fe II velocities (lower panel) from observations published by \cite{Terreran}. The model is shown with solid lines and the parameters used in the simulation are presented in Table~\ref{table:summary}.
\label{fig:LC1}} 
\end{figure}

We conducted first a exploratory analysis via $\chi^2$ minimization over the set of LC models presented in the previous section. From the derived tentative values, and based on the experience with other SNe, it was decided to vary the mass of the progenitor. Our preferred LC is presented in Figure~\ref{fig:LC1} and the model parameters are given in Table~\ref{table:summary}. This simulation was obtained with a main-sequence mass progenitor of $25 M_\odot$ which has a $\simeq 7~M_\odot$ He-rich core and $R\simeq 1200 R_\odot$ (see Appendix~\ref{appendix:progenitor} for details on the chemical abundances). The explosion was initiated by a thermal bomb that released an energy of 2~foe. 
We assumed $0.2$~$M_\odot$ of $^{56}$Ni and explored the parameters of the magnetar around the values obtained from our lower-mass models. As shown in Figure~\ref{fig:LC1} the LC data up to $\sim200$~days is reasonably well fitted by a magnetar with $t_p= 3$ days, and $E_{\rm rot}=0.8\times10^{51}$ erg. We assumed an interval of 20 days between the explosion and the first observed data. Note that we used the explosion date as a free parameter of the fit, only limited by the date of last non-detection, which occurred around 100~days before discovery \citep{Terreran}. 
For completeness, a comparison between our model photospheric velocities and the measured Fe~II line velocities is also presented in Figure~\ref{fig:LC1}. The model underestimates the iron velocities at early times, which could indicate that a different progenitor or a slightly more powerful magnetar could be required. However, note that although iron velocities are usually adopted as tracers of the photospheric velocity for normal SNe \citep[see e.g.][]{Takats2012}, this has not been fully proven for magnetar-powered SNe II.

\subsection{SN~2004em}
\label{sec:04em} 
Another interesting case is SN~2004em, the most extreme member of a small group of slowly rising Type~II SNe \citep{taddia16}.
\cite{Arcavi12} commented on the peculiar photometric behavior of SN~2004em. For the first few weeks it was similar to a Type II-P SN, while around day 25 it suddenly changed behavior to resemble a SN~1987A-like event, with similar long LC rise and expansion velocities. 
The total rise time was $\sim 110$~days, and only few additional observations were carried after the LC maximum.
Although it was not as bright as SLSNe, \cite{taddia16} modeled the LC with a rather large kinetic energy, $E_k=11.3$~foe and estimated $M_{\rm ej}\simeq 43 M_\odot$. Both kinetic energy and ejecta mass are the largest in their sample of long-rising SNe II, i.e. a rare family with only six members identified at the time\footnote{According to a later poster there were eight members, see {\tt http://sn2016.cl/documents/posters/poster\_taddia.pdf}}.
The extreme values of the physical parameters needed to model this object can be an indication that this SN may have been powered by other mechanisms. 
The radius of the progenitor and the degree of nickel mixing in \citet{taddia16} were derived from hydrodynamical modeling done with the SuperNova Explosion Code (SNEC) and based on progenitor stars constructed using MESA \citep{MESA2011} with radii of $320 - 350~R_\odot$ and nickel mixing of $25\%$ in the inner layers. \cite{taddia16} also inferred a value of $\Nimass \simeq 0.1~M_\odot$ from the tail of the LC.

\begin{figure}
\resizebox{\hsize}{!}{\includegraphics{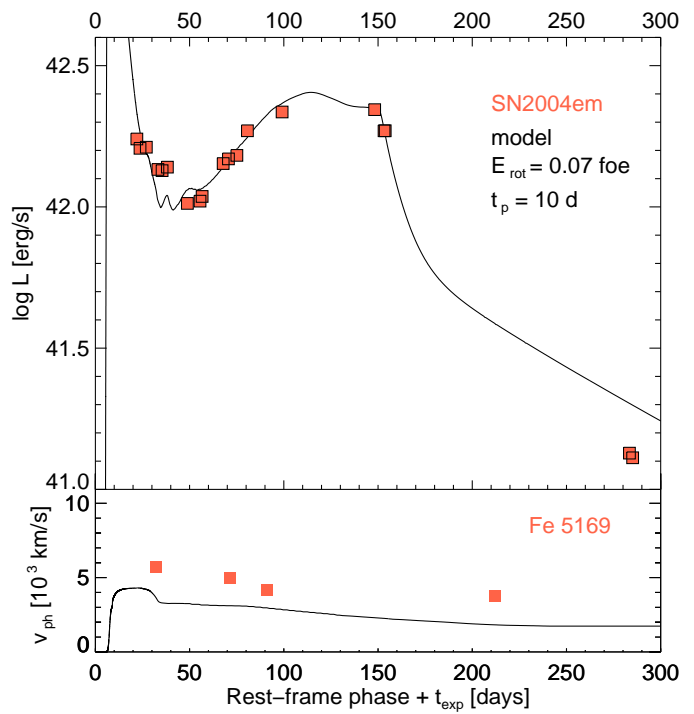}} 
\caption{Magnetar-powered model for SN~2004em. The SN is assumed to have exploded  
$t_{\rm exp}=10$~days before discovery. The points show the bolometric luminosities (upper panel) and Fe~II velocities (lower panel) from observations published by \cite{taddia16}. The model is shown with solid lines and the parameters used in the simulation are presented in Table~\ref{table:summary}. To reproduce the SN evolution during the cooling phase ($\sim$ first month) an interaction between the SN ejecta and some circumstellar material was assumed (see more details in \S~\ref{sec:04em}).
\label{fig:LC2}} 
\end{figure}

We performed a tentative fit to the LC of SN~2004em. Our main goal was to see if we can approximately reproduce the observed rise and $L_{\rm max}$ assuming a magnetar power source.
Our modeling procedure again started with a $\chi^2$ minimization using our grid of models for $15~M_\odot$ and $\Nimass \simeq 0.1~M_\odot$. The best fit was too bright, therefore we decreased the explosion energy to $E_k=0.8$~foe, and the magnetar parameters were adjusted to $E_{\rm rot}=0.07$~foe and $t_p=10$~days. With these parameters we could reproduce most of the LC (the slow rise plus broad maximum), however the observed decline during the early cooling phase was much more slower than the one shown by our models. 
In order to improve the match to the early LC we assumed that the star was surrounded by a diluted medium that is shocked by the SN ejecta. Such an interaction only modifies the early part of the LC. A good match with the data was obtained by assuming this CSM to be extended out to $\sim 3800~R_\odot$ and to contain a mass of $\approx 0.5 M_\odot$. This optimal model is shown in Figure~\ref{fig:LC2}. Although the choice of model parameters was based only on the LC, the photospheric velocity evolution compared to the Fe~II line velocities is shown for completeness in Figure~\ref{fig:LC2}. Similarly to the case of OGLE14-073, the model velocities underestimate the iron line velocities. 

\section{Comparison to other works}
\label{sec:comparison}
There are two recent works focused on the analysis of magnetar effects on type II SNe \citep{ST2017,DA17}. As here, the simulations were done using a one-dimensional radiation-hydrodynamics code assuming gray approximation for the radiation and including the magnetar source as an extra term in the energy equations assuming full energy trapping. The codes used in each work were different, as well as the initial setup. For example, \cite{DA17} (hereafter DA18) used an Eulerian code ({\sc heracles}) and they did not consider the radioactive decay. 

DA18 noticed that, in order to obtain density and temperature structures smooth at all times, an extended magnetar energy deposition is needed. Instead, here we have considered a limited range of mass where the  magnetar energy is deposited. similar to \cite{ST2017} prescription. The LCs of DA18 do not show a late time bump in the transition to the nebular phase. Such a bump is present in some of our cases, as well as in most of the LCs computed by \cite{ST2017}. 
Apart of this feature, the overall shapes of the LCs are similar. Note DA18 computed and discussed spectral features, while we made primary focus on the LCs. 
Regarding the fit to OGLE14-073, DA18 show two good matching models with ejected mass lower than ours ($M_{\rm ej} = 11.9\,M_\odot$ and $M_{\rm ej} = 17.8\,M_\odot$ versus $M_{\rm ej} \sim 20.5\;\,M_\odot$ in our case). They have obtained $E_{\rm rot}=0.4$~foe (versus $E_{\rm rot}=0.8$~foe) and $t_p=12$~d that is similar to our $t_p=10$~d. Given the degeneracies of the problem, all these results seems to be consistent.

\section{Discussion and conclusions} 
\label{sec:conclusion}

Magnetar-powered models generate a diversity of hydrogen-rich SNe: ordinary and brighter ones. We have explored a wide range of magnetar properties by varying their rotational energies $E_{\rm rot}$ and spindown timescales $t_p$. If the commonly accepted values for the inertia moment $I=1.3\times 10^{45}$~g~cm$^2$ and radius $R=10$~km are adopted, then the inversion of the expressions relating magnetar parameters are $P\approx \left(5/\sqrt{E_{\rm rot}}\right)$~ms, and $B\approx \left(1.25\times 10^{15}/\sqrt{E_{\rm rot} ~ t_p}\right)$~G; with the energy in units of foe and the spindown timescale in days.
This means that $t_p$ cannot be much smaller than our lowest value of $0.03$ days if we want to keep the magnetic field strength of the NS comparable to those of known magnetars \citep[see][for a reference about Galactic magnetars and their properties]{Olausen}. 

For a fixed progenitor mass of $15~M_\odot$ we found that magnetars spinning faster, but below physical breakup limit, produce more luminous events, being the spindown timescale related to the duration of the maximum brightness. Some combinations of the magnetar parameters produce a clear maximum in the LC followed by a smooth decline. In other cases, which we call intermediate LC morphologies, the declining slope breaks into a steeper tail.
A third case shows a similar LC to those of normal Type~II-P SNe (see the right panel of Fig~\ref{fig:cases}).
Very bright Type~II-P SNe are a distinctive class of events that have not been observed yet but can be produced by a magnetar source. A peculiar feature of these events is the existence of a phase when the luminosity increases by $\approx$ one order of magnitude before the plateau is settled. In addition, the numerical experiments performed here led us to propose that some peculiar SN~1987A-like SNe can be explained by the magnetar source. Interestingly, we were able to produce the slowly rising SN-1987A-like LC morphology without assuming the usual BSG structure. 
A summary of the magnetar and the stellar progenitor parameters is presented in Table~\ref{table:summary}, whereas the detailed chemical composition is given in Appendix~B. We note that both \Ni\ and magnetar energy depositions were taken into account in our calculations. They relative influence depends on the specific values of the parameters adopted, as shown by \cite{moriya2017}.

\begin{table*}
\begin{center}
  \caption[]{Summary of physical parameters for the SN progenitors used throughout this work. Values for the RSG configurations were obtained from stellar evolution calculations. The degree of \Ni\ mixing is given as a fraction of the interior mass of the model. The two rightmost columns provide the preferred magnetar parameters.}
\begin{tabular}{clcccccccccc}
\hline
\hline
\noalign{\smallskip}
          &$M_{\rm ZAMS}$ & $R$  & $M_{\rm Ni}$ & \Ni$_{\rm mix}$ & $X_{\rm sup}$ & $Y_{\rm sup}$ & $Z_{\rm sup}$ & $E_{\rm k}$[foe]& $E_{\rm rot}$ [foe]&$t_{\rm p}$ [d]\\
\noalign{\smallskip}\hline\noalign{\smallskip}
grid      & 15 $M_\odot$ & 500 $R_\odot$& 0.1 $M_\odot$ & 0.5    & 0.619 & 0.36  & 0.021  &  1.5   &\multicolumn{2}{c}{Table 1} \\
OGLE14-073& 25 $M_\odot$ &1200 $R_\odot$& 0.2 $M_\odot$ & 0.95   & 0.573 & 0.408 & 0.019  &  2.0         & 0.8  & 3 \\
SN2004em  & 15 $M_\odot$ $^*$ & 500 $R_\odot$$^*$& 0.1 $M_\odot$ & 0.5    & 0.619 & 0.36  & 0.021  &  0.8  & 0.07  & 10 \\
\noalign{\smallskip}
\hline\\[0.01cm]
\multicolumn{9}{l} {$^*$ Modified by adding $0.5 M_\odot$ of CSM extended out to 3800 $R_\odot$.}\cr
\end{tabular}	
  \label{table:summary}
\end{center}
\end{table*}

We have shown that magnetar-powered explosion models can explain the overall luminosity of two observed H-rich SNe: the recent interesting case of OGLE14-073 \citep{Terreran}, and the mildly bright SN~2004em \citep{taddia16}. Our preferred model for OGLE14-073 has $P\sim 5$~ms and $B\sim 7 \times 10^{14}$ G. For SN~2004em values of $P\sim19$~ms and $B\sim1.5\times 10^{15}$ G were found based on the LC modelling around maximum. However, the presence of some CSM was needed in order to reproduce the early observations. In both cases, an RSG progenitor was assumed, with $M_{\rm ZAMS}=$ 15 M$_\odot$ and  25 M$_\odot$ for SN~2004em and OGLE14-073, respectively.
The photospheric velocities of our models tend to lie below those measured from iron lines. Nevertheless, we were not focused on finding a model that reproduces both observables. Instead, our goal was to test whether a magnetar is able to reproduce the observed LC morphology of H-rich SNe. In any case, it is not clear if iron lines are an accurate tracer of the photospheric velocity in magnetar-powered objects, as is usually assumed for normal SNe II. Our models show that 1987A-like morphologies can be produced from RSG progenitors by including a magnetar source. 

During the nebular phase our treatment is too simplified to expect a reliable match with the observations. Among other reasons, because the spectral energy distribution of the magnetar is not specified, whereas the bolometric data derived from observations usually assume thermal emission. Here we have adopted a braking index of $n=3$ (defined from $\dot{\Omega}=-k\Omega^n$) from dipolar radiation \citep{Shapiro}, although a range of $1<n<2.8$ is observed in isolated pulsars. Therefore, a different exponent given by $-(n+1)/(n-1)$ is possible in the magnetar luminosity function (Eq.~\ref{magnetar}), which allows for a steeper decline.

We conclude that the observational appearance of SNe~II powered by magnetars is extremely varied. Future advances in the physics of this type of model will be very relevant. 

\begin{acknowledgements}
This work was partially supported by grant PI-UNRN2016-40B531. We thank to Omar Benvenuto for his valuable guidance in the development of the code and later modifications. We are grateful to G. Folatelli and to the anonimous referee for their help to improve the manuscript.
\end{acknowledgements}


\appendix\section{Equations of relativistic radiating hydrodynamics}\label{appendix:apendice_eqs}

The code we employed is a modified version of the one described in \cite{Bersten11}, and applications were already shown in \cite{Bersten16}. It is a one-dimensional Lagrangian 
code that solves explicitly hydrodynamic equations while it assumes an implicit strategy for energy conservation and flux limited radiative 
transport. Although it is usual to consider that Newtonian physics is adequate 
for computing light curves of supernovae, we find that powerful enough magnetars may force the expanding
envelopes to move at speeds that are a non negligible fraction of the speed of light (see e.g Fig 5 of Bersten et al. 2016). The physics assumed in our code needed a revision to properly handle relativistic velocities. 
For this purpose we have adopted the scheme presented by \cite{1979ApJ...232..558V}, who assumes that the object evolves adiabatically, an approximation
certainly not suitable for the problem we face in this paper. In order to include the corrections to the
radiative transfer together with conservation of energy we have adopted the description presented in \cite{1969qhea.conf..397M}. 
Here we detail the equations implemented in our SN light-curve code.

We assume a metric such as

\begin{equation}
ds^{2}= - e^{2\phi} c^{2} dt^{2} + \bigg( \frac{1}{\Gamma} \frac{\partial r}{\partial m} \bigg)^{2} dm^ {2} +  r^{2} d\Omega^{2},
\end{equation}

\noindent where $\Omega$ is the solid angle and

\begin{equation}
\Gamma^{2}= 1 + \bigg( \frac{U}{c} \bigg)^{2} - \frac{2G\tilde{m}}{rc^{2}}.
\end{equation}

\noindent The gravitational mass $\tilde{m}$ is given by

\begin{equation}
\tilde{m}(m)= \int_{0}^{m} (1+E/c^{2}) \Gamma dm^{'}.
\end{equation}

\noindent The velocity $U$ is

\begin{equation}
\frac{\partial r}{\partial t}= e^{\phi} D_{t} r= e^{\phi} U.
\end{equation}

\noindent The coefficient of the metric is given by

\begin{equation}
\frac{\partial \phi}{\partial P}= - \frac{V}{ w c^{2}},
\end{equation}

\noindent where $w$ is the relativistic enthalpy, given by

\begin{equation}
w= 1 + \frac{E+PV}{c^{2}}.
\end{equation}

\noindent At the stellar surface the coefficient of the metric is

\begin{equation}
e^{\phi_{s}}= \Gamma_{s}^{-1} \bigg( 1-\frac{2G\tilde{m_{s}}}{r_{s}c^{2}} \bigg). 
\end{equation}

\noindent The equation of motion of the fluid is

\begin{equation}
 \frac{\partial U}{\partial t}= e^{\phi} \bigg( 
 - \frac{4\pi\Gamma r^{2}}{w} \frac{\partial P}{\partial m}
 - \frac{G\tilde{m} }{r^{2}} 
 - \frac{4 \pi G r P}{c^{2}}.
 \bigg)
\end{equation}

\noindent The specific volume is

\begin{equation}
V= \frac{1}{\Gamma} \frac{\partial }{\partial m} \bigg( \frac{4 \pi}{3} r^{3} \bigg).
\end{equation}

\noindent The radiative luminosity is given by

\begin{equation}
L= - \big( 4 \pi r^{2} \big)^{2}\; \frac{ac}{3\kappa}\; e^{-4\phi}\; \frac{\partial }{\partial A} 
\bigg( e^{4\phi} T^{4} \bigg).
\end{equation}

\noindent Finally, the equation of energy conservation is

\begin{equation}
 D_{t} E + P  D_{t} V + e^{-2\phi}\; \frac{\partial }{\partial A} \bigg( e^{2\phi} L \bigg)= 0.
\end{equation}

\subsection{Results with the two solvers}\label{appendix:comparo}

\begin{figure}[h]
\centerline{\includegraphics[width=0.59\hsize,angle=0]{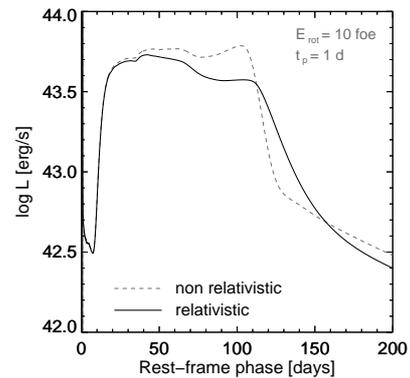}}
\caption{Light curve for the magnetar of Figure~\ref{fig:definition}.}
\label{fig:comparo1}
\end{figure}

\begin{figure}[h]
\resizebox{\hsize}{!}{\includegraphics[angle=-90]{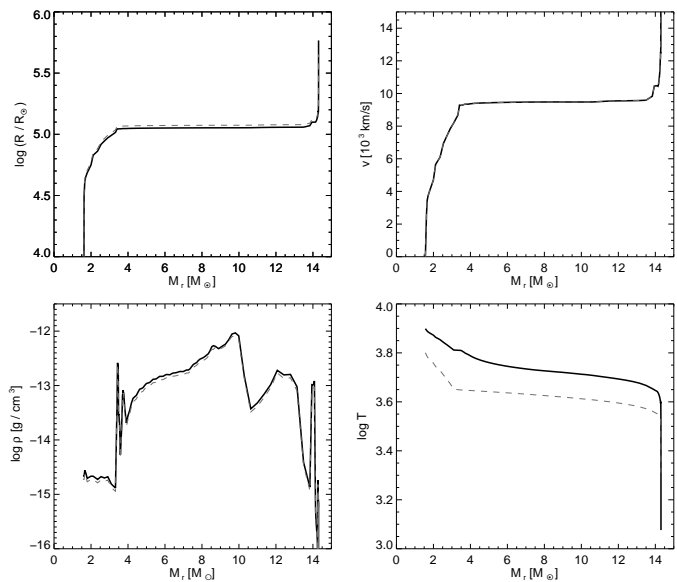}}
\caption{Inner physical quantities for the model with $E_{\rm rot}=10$~foe and $t_p=1$~d commented in detail in the text. These are profiles at time $t=104$ days, when the niquel phase is ending and the photosphere reciding. The continuous line corresponds to the relativistic solver and the dashed to the non relativistic. 
} 
\label{figure:comparo2}
\end{figure}

In standard core collapse explosion relativistic velocities are only relevant in the outermost layers of the progenitor. However, in presence of a very powerful magnetar, also  inner layers can acquire mildly relativistic velocities (e.g. as high as  6$\%$ of the speed of light in results presented by Bersten et al. 2016, whereas the outer layers reach $\sim 0.15\,c$).  
Here we compare the results obtained with  \cite{Bersten11} original treatment (plus magnetar) indicated as ``non relativistic" and with the present modified version denoted as ``relativistic". Figure~\ref{fig:comparo1} shows the LCs with both solvers for the model presented in Figure~\ref{fig:definition}. 
With this powerful magnetar the LC is modified as result of the incorporation of the relativistic radiating hydrodynamics, although the overall morphology, according to our simple scheme, is preserved. We should classify as plateau-kind the LCs resulting  with both solvers. However, a bump around $\sim 100$~d is less prominent with the relativistic treatment.
The profiles of density, radius, velocity and temperature for this model are presented in Figure~\ref{figure:comparo2}. The most clear difference is noted in the temperature profile.
If the temperature of the ejecta in the hydrogen rich layers changes around the temperature for hydrogen ionization this may substantially change the matter opacity and hence the outcoming luminosity.

%
\section{Progenitor chemical composition}\label{appendix:progenitor}

\begin{figure}
\resizebox{\hsize}{!}{\includegraphics[angle=-90]{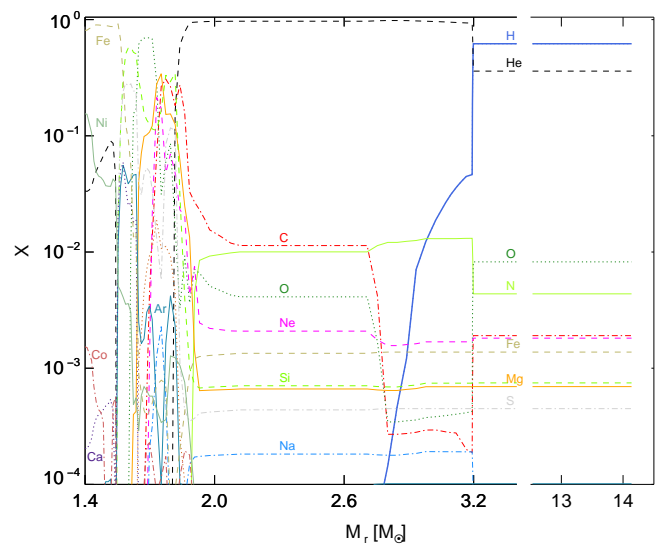}}
\caption{Chemical composition of a $M_{\rm ZAMS}=15M_\odot$ star during the RSG state.} 
\label{X_M15}
\end{figure}

\begin{figure}
\resizebox{\hsize}{!}{\includegraphics[angle=-90]{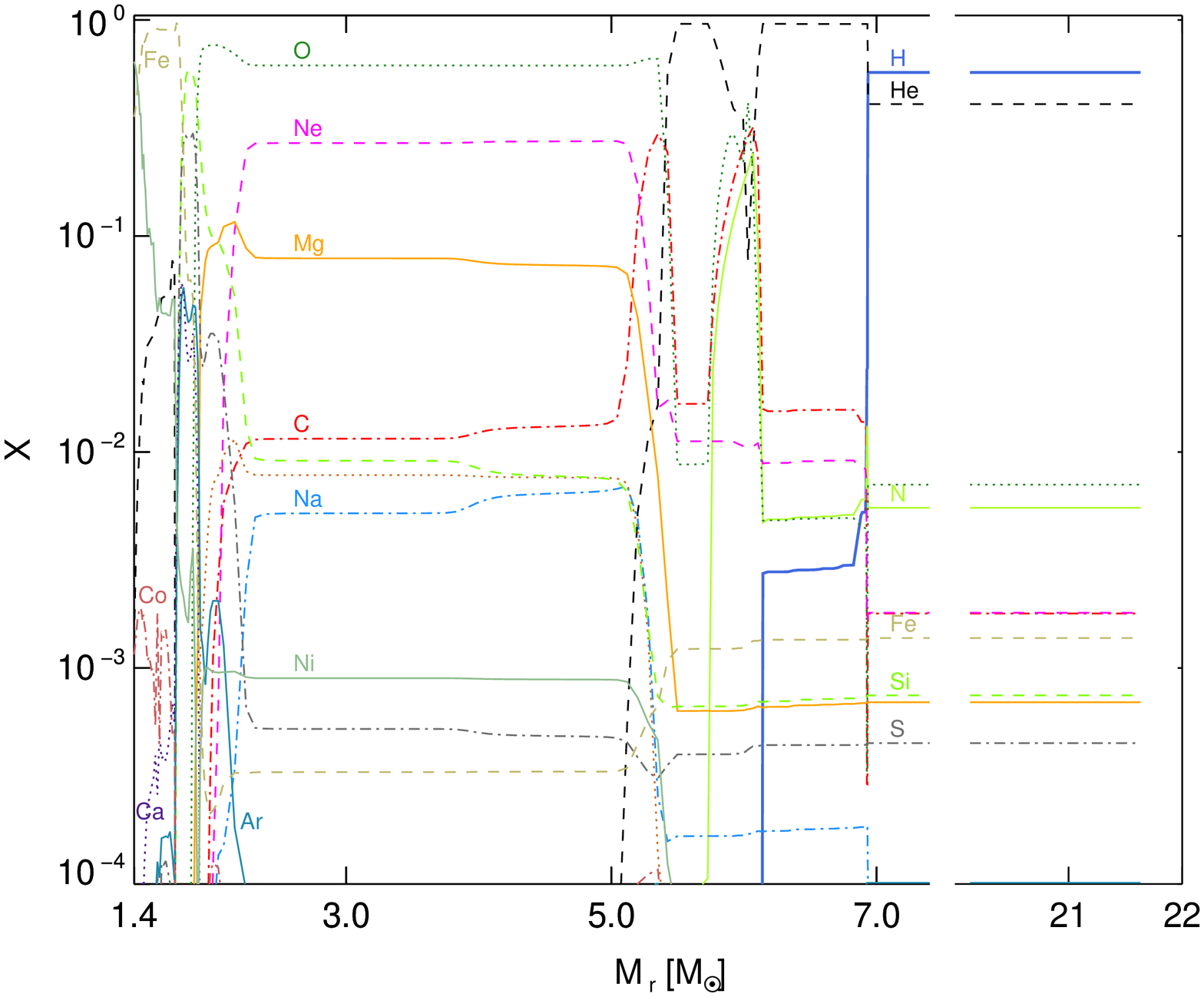}}
\caption{Chemical composition of a $M_{\rm ZAMS}=25M_\odot$ star during the RSG state.} 
\label{X_M25}
\end{figure}

We have considered RSG structures calculated by \cite{NH88} as our SN progenitors. For completeness we provide their detailed composition in Figures~\ref{X_M15} and \ref{X_M25} for models with main-sequence masses of 15 and 25 M${_\odot}$ respectively. Note that the internal core is removed for simplicity as it is considered to collapse and to form the magnetar. Chemical stratification presented is the result of stellar evolution calculations. However, as in other studies the \Ni~distribution (modified by chemical mixing) was adapted for convenience and assuming a conservative value of $0.1 M_\odot$ in our grid calculations.

\end{document}